%% file: Tinder.tex
\documentclass{IEEEtran}

\usepackage{hyperref}
\usepackage{url}
\usepackage{graphicx}
\usepackage{subfigure}
\usepackage{times}
\usepackage{balance}
\usepackage{xcolor}
\usepackage{amsmath}
\usepackage{balance}
\usepackage{xspace}
\usepackage{epstopdf}

\newcommand{\one}{({\em i}\/)}
\newcommand{\two}{({\em ii}\/)}
\newcommand{\three}{({\em iii}\/)}
\newcommand{\four}{({\em iv}\/)}
\newcommand{\five}{({\em v}\/)}

\def\eg{\emph{e.g.,}\xspace}
\def\etc{\emph{etc.}\xspace}
\def\ie{\emph{i.e.,}\xspace}
\def\etal{\emph{et al.}\xspace}

\begin{document}

\setlength{\pdfpagewidth}{8.5in}
\setlength{\pdfpageheight}{11in}

\title{A First Look at User Activity on Tinder}

%

\author{Gareth Tyson,$^1$ Vasile C.\ Perta,$^2$ Hamed Haddadi,$^{1}$ Michael C.\ Seto,$^3$ \\
$^1$Queen Mary University of London, $^2$Sapienza University of Rome, $^3$Royal Ottawa Health Care Group}

%
%

\maketitle

\begin{abstract}

Mobile dating apps have become a popular means to meet potential partners. Although several exist, one recent addition stands out amongst all others. \emph{Tinder} presents its users with pictures of people geographically nearby, whom they can either \emph{like} or \emph{dislike} based on first impressions. If two users like each other, they are allowed to initiate a conversation via the chat feature. In this paper we use a set of curated profiles to explore the behaviour of men and women in Tinder. We reveal differences between the way men and women interact with the app, highlighting the strategies employed. Women attain large numbers of matches rapidly, whilst men only slowly accumulate matches.
To expand on our findings, we collect survey data to understand user intentions on Tinder. 
Most notably, our results indicate that a little effort in grooming profiles, especially for male users, goes a long way in attracting attention.
 
\end{abstract}

\section{Introduction}


Online dating has become extremely popular, with 38\% of American adults who are ``single or looking'' having experimented with it~\cite{smith2013online}. Mobile dating apps have become particularly prevalent in recent years~\cite{valkenburg2007visits}. The most noticeable shift these apps enable is the ability to discover and interact with nearby potential mates. Location-based services have long been touted as a commercial revolution (\eg hailing taxis). However, only recently has the discovery of people become popular. One app stands out in this regard. \textit{Tinder} presents users with pictures of other nearby people; users are then allowed to either ``\emph{like}'' or ``\emph{dislike}'' the picture(s). If two users like each other, they will be given the opportunity to interact via text messaging: this is called a \emph{match}. Whereas previous dating services have aimed to match on interests, Tinder, instead, matches on locality. Thus, by focusing on first impressions, Tinder constitutes a cut-down version of online dating, without any of the features that make it possible to understand the deeper characteristics of potential mates. 


Despite Tinder's popularity, there is a distinct lack of research on the app's usage. As a primarily heterosexual dating app, we are particularly curious to understand the interactions between the two genders, as its novel style of matching and ambiguity in user intent raises many questions that have not been explored yet. In this paper, we pursue two key avenues of study. First, we ask how gender impacts matching and messaging rates for Tinder profiles. Past work has established clear differences between how genders pursue romantic engagement; as such, we are curious to understand how this translates to the novel matching and usage style of Tinder. Second, we ask, what profile characteristics are common in Tinder, as well as which characteristics can impact matching and messaging rates. Although past online dating research has shed insight on user behaviour~\cite{hitsch2010makes}, these traditional services differ greatly from how Tinder matches users.





To explore these questions, we have performed a measurement campaign of Tinder. We have created a number of curated profiles~\cite{webb2008social}, which we have injected into London and New York (\S\ref{sec:dataset}). We have used these profiles to monitor the way others react to them, specifically in terms of matches and subsequent messaging. Through data on almost half a million users, we show that the two genders exhibit quite different matching and messaging trends (\S\ref{sec:matching}). Women tend to be highly selective in whom they like, leading to a starvation of matches for men. Men, on the other hand, are more accommodating in their practices, hitting like for a far larger proportion of women. This mirrors many sociological observations about mating, although Tinder seems to enact quite extreme examples of this. Our findings suggest a ``feedback loop'', whereby men are driven to be less selective in the hope of attaining a match, whilst women are increasingly driven to be more selective, safe in the knowledge that any profiles they like will probably result in a match. This leads us to explore key characteristics that impact user matching rates (\S\ref{sec:genders}). We show that simple improvements (\eg including more pictures) can substantially increase popularity, particularly for men. Interestingly, these increases come primarily from women who react more strongly to these profile improvements. In addition, we also perform a user questionnaire to validate and expand our interpretations of the data (\S\ref{sec:survey}).
\section{Background}
\label{sec:background}

\subsection{A brief tour of Tinder}

Tinder is a mobile dating app launched in 2012. Tinder profiles are very limited, containing just a name, age, interests and a short bio. Users stipulate their desired match by selecting the age and gender range, as well as writing a short description of themselves. When a user turns on the app, their location is reported to Tinder's server, which then returns a set of profiles matching the user's stipulated criteria within a given range (maximum is 100 miles). The user is then presented with a picture of a nearby user. This screen contains two large buttons, labelled with a cross and a heart. These allow the user to stipulate if they like (termed ``swiping right'') or dislike the profile (``swiping left''). If two users say that they like each other, they each are notified (otherwise the two users are not notified of anything). From this point forward, the two users can interact via text message within the app. This is the limit of the app's functionality and, as such, it constitutes an extremely cut-down version of an online dating experience. In fact, there is no formal means of reporting what a user desires from a match and, therefore, Tinder can even be used for simply meeting new friends.


\subsection{Related work}

Despite Tinder's growing popularity and unconventional matching style, it has received limited attention from the research community. Most related is a recent study into the privacy of mobile dating apps like Tinder~\cite{toch2013locality}. They observed a range of potential privacy concerns, primarily relating to the ability of attackers to track user location. This is perhaps exacerbated by the frequent use of location-based dating apps for immediate sexual encounters~\cite{Handel12}. Beyond this, little is known about the nature and usage of Tinder. That said, there is a significant body of research looking more broadly at online dating services that match people based on interests.

One of the largest studies was performed by Rudder~\cite{rudder2014dataclysm}. A number of interesting insights were gained, building a model of pictures that others are attracted to. These findings are backed by decades of psychology research, showing that initial physical attractiveness is often associated with other positive attributes, \eg intelligence~\cite{langlois2000maxims}. Although laboratory research provides greater control over study conditions, Tinder provides interesting potential, as it can provide data on in-the-wild first impressions on physical attractiveness. Fiore \etal also found that free-text components can play an important role in predicting attractiveness~\cite{fiore2008assessing,fiore2010}. In Tinder, this latter component is heavily suppressed, favouring first impressions based on pictures. There have also been a number of studies into user interaction on dating websites. For instance, it has been found that people use a variety of strategies to reduce uncertainty when interacting with new people~\cite{tidwell2002computer}. Disclosure of information is an important part of this~\cite{gibbs2010first}, as deception has been found to be commonplace~\cite{hancock2007truth}. This issue is particularly important in Tinder, as initial disclosure of information is very limited (due to the simplicity of profiles), although past work has found that direct conversation is much more important than preferences stipulated through profile bios~\cite{akehurst2012explicit}.

%
A key focus of this paper is gender. Several prominent sociological studies have investigated the nature of online dating across genders~\cite{toma2010looks,whitty2004cyber}, as well as the characteristics of the people using such services~\cite{valkenburg2007visits}. For example, Hitsch~\etal found that, on average, women show greater preference than men for income over physical attractiveness~\cite{hitsch2010makes}. This is consistent with more traditional social theory, which has found that, on average, men pay more attention than women to youth and physical attractiveness, whilst women place more prominence on social status~\cite{lance1998gender,Fisman06}. Interestingly, the cut down nature of Tinder means that users are forced into making decisions on primarily aesthetic qualities. Bolig \etal found that personal adverts left in a magazine's lonely hearts column tended to show that  physical characteristics are most salient in identifying potential mates~\cite{bolig1984self}. This is powerful motivation for an app like Tinder, however, it is unclear how textual descriptions of one's self in a magazine compare to pictures in Tinder. Underpinning this, various works have investigated what is considered attractive, such as a female preference for taller men~\cite{hitsch2010makes}, and a male preference for larger eyes~\cite{rhodes2006evolutionary}.
Perhaps most profound is the tendency towards homophily for users of online dating services, \ie the propensity to pursue partners who are similar to themselves~\cite{fiore2005homophily}. Interestingly, however, Hitsch \etal noted that in online settings, users tend to break away from traditional models of homophily measured by attractiveness; instead, users pursue attractive users regardless of their own appearance. Currently, how this maps to Tinder is unknown. We are particularly curious to see how this pursuit translates into matching and messaging trends between men and women. For instance, it has been found that, in traditional dating websites, men and women exhibit similar messaging rates early on in their lifecycles, but women tend to reduce the number of messages sent over time while men continue at the same rate~\cite{xia2013study}.



In this paper, we provide vantage on Tinder. As much as possible, unlike the above studies, we avoid categorising users into groups of attractiveness or desirability. Instead, we simplify our analysis by solely investigating the differences between male and female profiles. It has been long understood that the genders exhibit quite different patterns of behaviour in dating. However, in contrast to prior studies, we posit that Tinder suffers from far greater ambiguity; whereas other dating services make a particular effort to pair appropriate people, Tinder makes no such allowances. There is even a lack of guidelines that might create a commonality of intent between the user base. Instead, Tinder's purpose is left open to interpretation, allowing the ecosystem to emerge from a bottom-up perspective. Thus, profiling its usage is key for understanding this recent social phenomenon. To the best of our knowledge, this is the first measurement study of Tinder.

\section{Methodology} \label{sec:dataset}

\subsection{Data Collection}

To study Tinder we have performed a measurement campaign similar to~\cite{webb2008social}. This has involved injecting new profiles specifically designed to record interactions initiated by other users. This allows us to \one~collect thousands of user profiles, including the demographics, bios \etc; and \two~see which of our profiles gain the most likes from other users. Our methodology involves two stages: \one~manually creating curated profiles; and \two~ injecting the profiles into a locale to collect data. Table~\ref{tab:data_overview} presents an overview of the 14 curated profiles we created. To design these profiles, we have followed the methodologies of past studies~\cite{webb2008social,haddadi2010add}. We use simple profiles that reflect the characteristics of an ``average'' user, as defined by preliminary measurements. Our methodology is intended to reduce the number of variables that could potentially conflate results. With this in mind, we restrict  profile pictures to a single facial shot and do not include any biography. We use facial shots to prevent users interpreting extended information from the clothing or background setting, \eg income, education~\cite{langlois2000maxims}. We also exclusively use Caucasian pictures to avoid the complexities introduced by racial homophily~\cite{fiore2005homophily}.

We next describe the pictures used for the 14 profiles described in Table~\ref{tab:data_overview}.\footnote{Pictures are available at: \url{http://www.eecs.qmul.ac.uk/~tysong/tinder/pics.html}}
All profiles (bar 2) were placed in London, to remove the bias introduced by different cities. The profiles entitled \emph{stock} are generated using a set of copyright-free stock photos (each profile has one picture). The profiles entitled \emph{no-pic} contain no pictures, whereas the \emph{account-disabled} photos contain a picture saying the account has been disabled.
These were used as a benchmark against which the picture-enabled profiles can be compared. Finally, we also created profiles of a male and a female volunteer to allow us to request extra pictures; these were placed in New York to avoid bias caused by potential real-world relationships in London (note, we only compare across results for profiles in the same city). In all cases, we set the profile age to 24 as this was the most frequently observed age in our early data.

Once all profiles had been created, we wrote software to automatically register the profiles as available in a given location. The software then exhaustively retrieved all profiles within a 100 mile radius and clicked like for each one. If any profiles generated a match, we recorded the timestamps of the match and messages sent. Metadata (\ie age, bio, number of pictures) was recorded for all profiles returned. It is worth noting that our female profiles performed fewer likes than their male counterparts. This occurred because Tinder automatically limits users with excessive numbers of matches (our female profiles accumulated matches faster than our males, \emph{c.f.,} \S\ref{sec:matching}). To mitigate the impact of this, we separate all analysis into male and female groups, ensuring only \emph{proportional} comparisons are performed between groups.

In total, we collected 230k male profiles, and 250k female profiles. 12\% of male profiles were homosexual or bisexual, whereas this was the case for only 0.01\% of female profiles.\footnote{We categorise a user as homosexual/bisexual if the profile is returned to one of our profiles of the same gender.} As such, we focus primarily on heterosexual users. We emphasise that our study is \emph{not} intended to measure attributes like beauty or attraction. Instead, the above profiles are intended to provide us with a vantage point into Tinder. We therefore solely use them to trace behaviour, rather than attempting to measure fine-grained concepts of attractiveness. We later explore the importance of this distinction.

\begin{table}[t]
\center
\begin{tabular}{l | c | c  }
\hline
Profile        &            \#Likes   &       \#Matches   \\
\hline\hline
male-stock-1        &        43290   &        238     \\
male-stock-2          &      44543     &      309     \\
male-stock-3         &       43255      &     338    \\
female-stock-1       &       6631       &     682       \\
female-stock-2        &      10550     &      1109   \\
female-stock-3         &     5652       &     577    \\
\hline
male-no-pic           &      47695       &    79       \\
female-no-pic    &     9554       &     610     \\
male-account-disabled    &   71207     &      109     \\
female-account-disabled  &   10526      &     936   \\
\hline
male-real-1pic         &   86440      &     234    \\
male-real-3pic       &     79936      &     1568       \\
female-real-1pic       &     9337         &   1681    \\
female-real-3pic      &      10030      &     2319    \\
\hline
\end{tabular}
\caption{Overview of measurement profiles}
\label{tab:data_overview}
\end{table}

%
%

\subsection{Ethical Considerations} \label{sec:ethics}

We have gone through Institutional Review Board procedures. During this process, we made a number of considerations. A first key concern was that personally identifiable information may be revealed during the data collection.
We therefore avoided recording user profile pictures or names. Further, we did not collect message data sent by users. To analyse user bios, it was necessary to collect the text data. However, such data is publicly available and, as such, users are already aware that unknown third parties will be accessing it. The second concern was the nature of participants' interactions. Thousands of users matched with our curated profiles, some of whom sent messages. In all cases, we did not respond. This wasted the time of users and potentially generated frustration. However, a critical fact is that Tinder is highly ephemeral with few parallels to traditional dating services. Users may `like' hundreds of profiles per day and our questionnaire revealed that not initiating conversations and/or meet-ups are \emph{very} common. Consequently, liking one of our curated profiles will generate little disruption for users. As such, due to the game-like nature of the process, the impact is trivial; this is an observation we made through consultation with many real Tinder users.


\section{A match made in heaven?} \label{sec:matching}


\begin{figure}[t]
\center
\includegraphics[width=.50\textwidth]{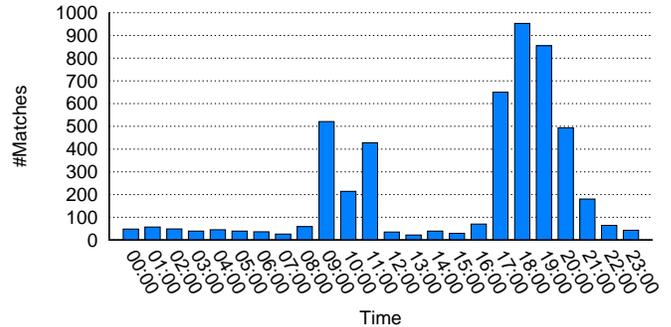}
\caption{Match arrival distribution}
\label{fig:match_arrival}
\end{figure}

First, we wish to explore the differences in matching and messaging rates for male and female users on Tinder. We hypothesise that, in-line with evolutionary social theory, gender will have a significant impact on these measures. We begin by simply inspecting \emph{when} users choose to use Tinder. Figure~\ref{fig:match_arrival} presents the matching across the time of the day for our London profiles. Clear diurnal patterns can be seen. Although activity is observed across the entire day, users tend to peak around 9:00 and 18:00. These are prime commuting hours in London; clearly, users have a tendency to use Tinder to pass the time during their commute. This is just one of the many benefits (or side effects) of embedding dating into mobile devices. Usage also continues into the evening, with matches reducing in the later hours (21:00 onwards). These patterns are consistent across both male and female users.

\begin{figure}[t]
\center
\includegraphics[width=.50\textwidth]{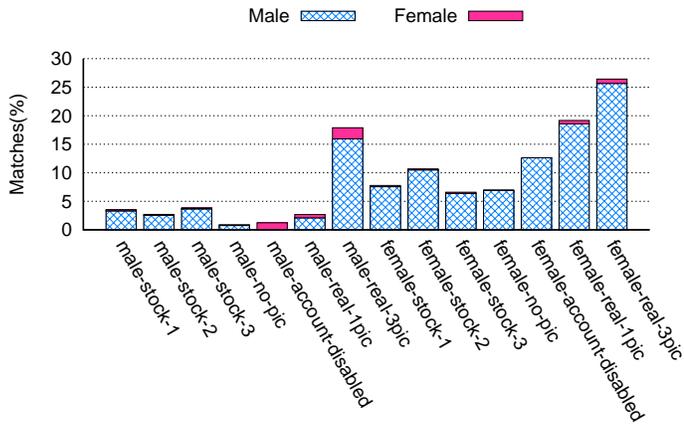}
\vspace{-16pt}
\caption{The number of matches per profile}
\label{fig:num_matches}
\end{figure}

\begin{figure}[t]
\center
\includegraphics[width=.42\textwidth]{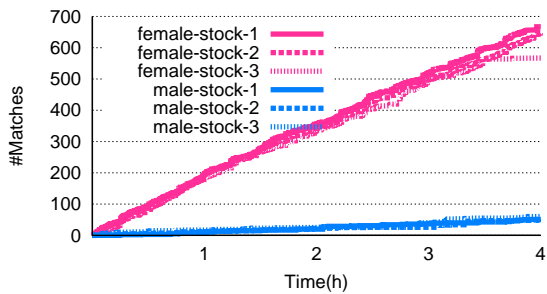}
\caption{The number of matches over time per profile (only stock profiles)}
\label{fig:matches_time_stock}
\end{figure}

Next, we ask if there is a noticeable difference in the popularity of our male and female profiles, as measured by matching and messaging rates. Figure~\ref{fig:num_matches} shows the percentage of matches obtained across our various stock profiles. This confirms a stark contrast. Our male profiles like a large number of other users, but only match with a small minority (0.6\%). The opposite can be seen for our female profiles, who attain a far higher matching rate (10.5\%). Of course, we cannot be sure that \emph{every} user we liked was also presented with our profile and, as such, these values offer a lower-bound on \emph{potential} matches. That said, the contrast between our male and female profiles is stark. More remarkable is the observation that nearly all matches received come from men. This includes our male profiles, indicating that homosexual men are far more active in liking than heterosexual women. Even though the male:female ratio in our dataset is roughly even, on average, 86\% of all the matches our male profiles receive come from other men. There is, however, a notable outlier. Our male-account-disabled profile actually acquires all of its matches from females. This profile simply has a picture stating that Tinder has taken down the account. 


These observations are also reflected in the temporal trends of matching. Figure~\ref{fig:matches_time_stock} shows how the matches occur over time with our stock photo profiles. Quite distinct trends are present that are common amongst all profiles belonging to each gender. Male profiles slowly build up matches over time, with a very shallow gradient of increase. In contrast, female profiles gain rapid popularity, achieving in excess of 200 matches in the first hour. The linear nature of these likes is particularly curious with a constant probability of matching over time. As such, algorithmic techniques for facilitating earlier matches would be highly beneficial for men, but largely unnecessary for women.




Once a profile has matched, the two users can exchange messages.  Again, differences can be seen. Overall, we find that 21\% of female matches send a message, whereas only 7\% of male matches send a message. Thus, women who match with us are 3 times more engaged than men. This is likely driven by the sheer number of male matches. Overall, we received 8248 male matches, most of whom do not pursue interaction. In contrast, we garnered only 532 female matches, suggesting that they are more careful with whom they like and therefore consider it more worthwhile to send a message~\cite{trivers1972parental}. This is rather different to other online dating services, where messages are usually the initial means of establishing interaction (without the prior need to ``match'').


\begin{figure}[t]
\center
\includegraphics[width=.42\textwidth]{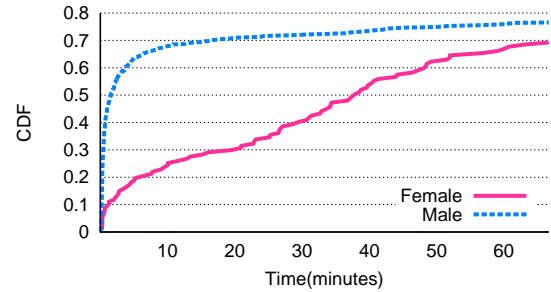}
\caption{Delay between match and first message sent}
\label{fig:messages_delay}
\end{figure}

\begin{figure}[t]
\center
\includegraphics[width=.42\textwidth]{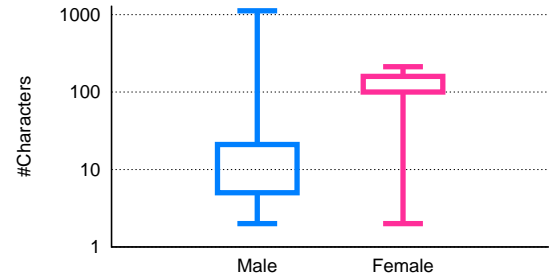}
\caption{Distribution of message lengths}
\label{fig:message_length}
\end{figure}

A further difference is the speed by which users pursue interaction.  Figure~\ref{fig:messages_delay} presents the cumulative distribution function (CDF) of the delay between a match occurring and a message being sent. This reveals a significantly faster pace of interaction than that seen in traditional online dating~\cite{fiore2010}. 63\% of messages sent by men occur within 5 minutes of the match taking place. This is only 18\% for women, suggesting that female users often wait to receive a message first. The median delay for sending messages is just 2 minutes for men, compared to 38 minutes for women. This could be driven by several factors, but it is well known that men often have to compete and differentiate themselves more as part of the mating ritual~\cite{kruglanski2012handbook}. Their efforts, however, are not always particularly emphatic. Figure~\ref{fig:message_length} shows the message length distribution. The median message length sent by men is only 12 characters, compared to 122 from women. For men, 25\% of message are under 6 characters (presumably ``hello'' or ``hi''). Consequently, it is clear that little information is being imparted in opening conversations.

\begin{figure}[t]
\center
\includegraphics[width=.42\textwidth]{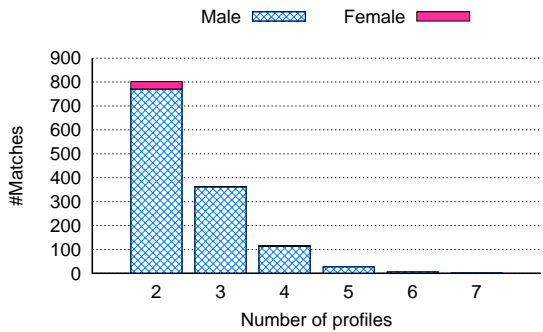}
\caption{Number of our profiles other users have matched with}
\label{fig:overlap}
\end{figure}

The above results show that men are willing to like a larger proportion of women. In the most extreme case, this could involve clicking like for all users encountered. In contrast, women are far more discerning. We conjecture that this creates a ``feedback loop'' in Tinder. Men see that they are matching with few people, and therefore become even less discerning; women, on the other hand, find that they match with most men, and therefore become even more discerning. To briefly explore this hypothesis, Figure~\ref{fig:overlap} presents the number of our profiles that each individual user matched with. Only 6\% of women liked multiple of our accounts, whilst 16\% of male profiles match with multiples of our profiles. This confirms that men are more liberal in whom they like. 4\% of male profiles even match with in excess of three of our profiles. 


\section{A Tale of Two Genders} \label{sec:genders}


The previous section has measured the differing popularities of our male and female profiles. We next take an exploratory approach, characterising all Tinder profiles collected by their age, profile pictures and bio. To investigate the importance of these three attributes we also perform controlled experiments to observe their impact on popularity.

\subsection{Age}

\begin{figure}[t]
 \center
 \includegraphics[width=.42\textwidth]{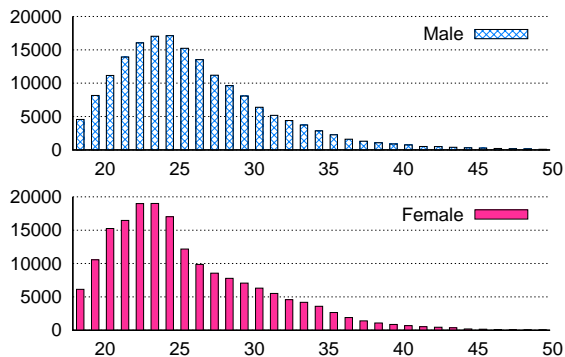}
 \caption{Age distribution of male and female users}
  \label{fig:demo:age}
\end{figure}

\begin{figure}[t]
 \center
 \includegraphics[width=.42\textwidth]{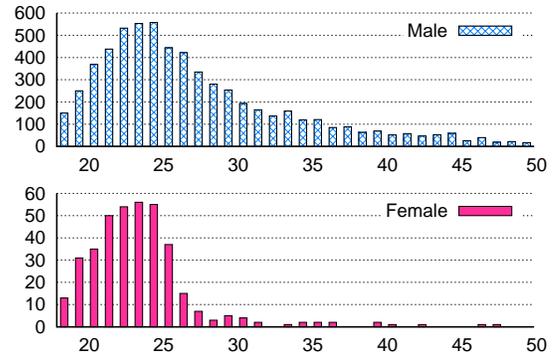}
 \caption{Age distribution of $(a)$ male users matching with our female profiles and $(b)$ female users matching with our male profiles}
  \label{fig:demo:age_match_male_accounts}
\end{figure}

We begin by inspecting the age distribution of users. Figure~\ref{fig:demo:age} presents a histogram of the age distribution for both male and female profiles. There are broad similarities across the two genders, although women tend to be marginally younger. The mean age for women is 25.2, compared to 25.7 for men. This is somewhat different to more typical online dating platforms, where there is a greater prominence of older women compared to men~\cite{xia2013study}. This is potentially due to the reputation that Tinder has for facilitating immediate and short-term relationships, which is often less desired by older women~\cite{kruglanski2012handbook}. We can also compare this age distribution with the matches received for our profiles (all of which were set to 24 years old). Figures~\ref{fig:demo:age_match_male_accounts} plots the ages of other users matching with our profiles. It can be seen that the distribution of male users matching with us is consistent with the overall population (an average of 25.7 vs.\ 25.8), whereas it is different for female profiles. Across the whole dataset, the average age of females is 25.2; for matches, this drops to 24.3. This is caused by a lack of older women matching with our 24 year old profile; a propensity that is frequently observed in mating~\cite{kruglanski2012handbook}. It is also mirrored in other online dating environments; Fiore \etal found that women in their 20s and 30s were more likely to interact with older men, whereas men consistently sought younger women~\cite{fiore2010}.







\subsection{Profile pictures}

\begin{figure}[t]
 \center
 \includegraphics[width=.38\textwidth]{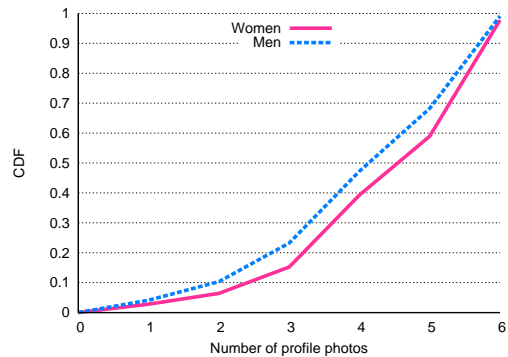}
 \caption{CDF of number of pictures on male and female profiles}
  \label{fig:demo:num_pics}
\end{figure}

It has been found that 77\% of online dating profile views are for profiles with at least one photo~\cite{hitsch2010makes}. Hence, we next explore the role of profile pictures in Tinder. Figure~\ref{fig:demo:num_pics} presents the CDF of the number of profile pictures for males and female profiles. This reveals a relatively similar numbers of pictures on male and female profiles: an average of 4.4 for men vs.\ 4.9 for women. The next question is therefore how do the genders view the importance of the number of pictures?

\begin{figure}[t]
\center
\includegraphics[width=.42\textwidth]{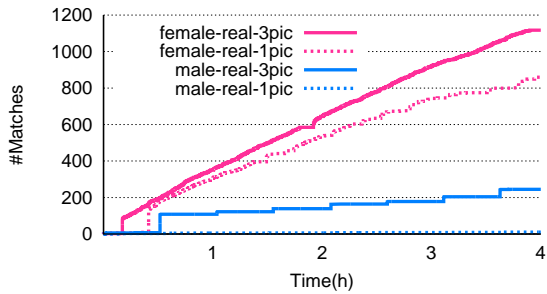}
\caption{The number of matches over time per profile (only real profiles)}
\label{fig:matches_time_real}
\end{figure}

To explore this, Figure~\ref{fig:matches_time_real} shows the number of matches over time for our two \emph{real} profiles (see Table~\ref{tab:data_overview}) when using differing numbers of profile pictures. Note that our \emph{real} profiles were those contributed by volunteers to allow us to request multiple specific photos (see \S\ref{sec:dataset}). It can be seen that the number of pictures has a notable impact. When increasing the female profile from 1 to 3 pictures, a 37\% increase is seen in matches. This is even more significant in male profiles. With a single profile picture, after 4 hours, only 44 matches were made, whereas this increased to 238 with three pictures. Even more interesting is the observation that the fraction of women liking our male profile increases most dramatically. After four hours, only 14 of our male profile matches were from women; with three pictures, this number increased to 65. This is likely partly driven by the greater concern that women have of deception~\cite{gibbs2010first}, alongside their preference for deeper information about mates~\cite{trivers1972parental}. Either way, the need for men to have multiple pictures is far greater than for women. Those attempting to improve their matching rates should consider this closely.

\begin{figure}[t]
\center
\includegraphics[width=.50\textwidth]{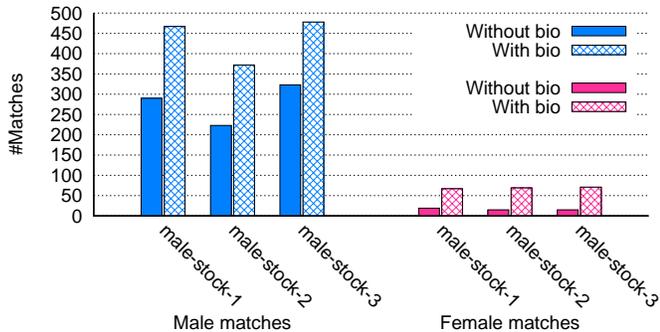}
\caption{Matches per profile: bio vs non-bio (only male stock profiles) }
\label{fig:matches_bio_vs_nobio}
\end{figure}

\subsection{Bios}

\begin{figure}[t]
\center
\includegraphics[width=.42\textwidth]{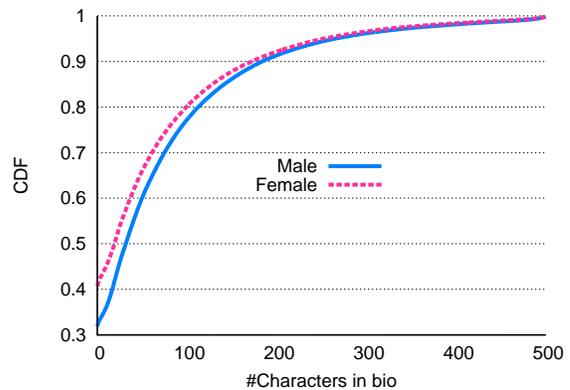}
\caption{CDF of the number of characters in user bios}
\label{fig:bio_length}
\end{figure}

Tinder allows users to provide a short biography about themselves. Figure~\ref{fig:bio_length} presents the distribution of bio lengths (number of characters) for both male and female profiles. In-line with the core principles of Tinder, people do not provide much information. 36\% of accounts have no bio, with the majority under 100 characters (maximum is 500). This is particularly prevalent amongst women, 42\% of whom have blank bios. Considering that free-text within dating profiles can heavily contribute to attractiveness~\cite{fiore2008assessing}, this tendency in Tinder could be seen as a negative one. Likely the ability of female profiles to gain high matching rates regardless has reduced the value of text. Their perceived importance is greater for male profiles though~\cite{kruglanski2012handbook}, shown by the smaller proportion of male profiles with blank bios (30\%). 

To investigate this hypthoesis we recreated our male stock profiles, but with short bios simply saying hello and that they are from London. We focus on male profiles as female profiles do well regardless. Figure~\ref{fig:matches_bio_vs_nobio} presents the matches attained. For each of our male profiles, we present the number of matches they attain both with and without a bio. The blue bars count the number of matches accumulated from other men, whilst the pink bars count the number of matches accumulated from other women. In all cases, the profiles with bios do far better. This is particularly the case for acquiring female matches. Without bios, our male stock profiles received an average of 16 matches from women; this increases four-fold to 69 with a bio. The number of matches from men also increases, but far less substantially (by 58\% on average).

\section{Exploring User Perspectives} \label{sec:survey}

\input{stats.tex}

We next seek to validate and further explore the meaning behind our findings. To achieve this, we have performed a user questionnaire. This is intended to provide context to our earlier findings. The questionnaire was distributed via social media and various mailing lists. Naturally, self-selection and reporting could introduce bias and, thus, we use the survey data primarily to drive discussion. 
Due to a lack of data on homosexual users, we filter all entries to leave users seeking heterosexual interactions; this leaves \nresponses responses, \nmales from men, and \nfemales from women. In-line with our earlier empirical observations, nearly all participants (\punderthirtyfive\%) were under the age of 35. We primarily recruited frequent users, with \puseonceaweek\% reporting using Tinder at least once a week. Interestingly, we also noted that many users (\ponlinedatedbefore\%) stated they had used other online dating services before, indicating that Tinder is not necessarily the first (or only) port of call for new online daters.

\begin{figure}[t]
 \center
 \includegraphics[width=80mm]{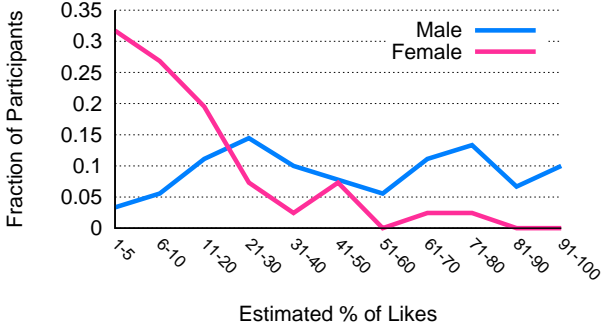}
 \caption{Probability density of estimated percentage of likes}
  \label{fig:survey:estimate_likes}
\end{figure}

To support our earlier findings, we began by asking users to estimate the percentage of profiles that they click like for. Figure~\ref{fig:survey:estimate_likes} presents the probability density function of the responses for the two genders. In-line with expectations, women report that they are far more selective in whom they like: \pwomenlikeunder11\% of women estimate that they like 10\% or fewer of all profiles they encounter. In contrast, only \pmenlikeunder11\% of men report such high selectivity. To explore what led to this, we went on to ask participants what their intentions were when using Tinder. Specifically, we asked \textit{``Please rank your intentions when using Tinder''}. We gave several options in the questionnaire:
\one~I use Tinder to look at profiles; 
\two~I use Tinder to chat with people online; 
\three~I use Tinder to find a partner; 
\four~I use Tinder for casual dating; and 
\five~I use Tinder for one night stands. 
We asked the participants to rank each option from 1 to 5 (with 5 being the most important). Table~\ref{tab:survey:intent} presents the mean score allocated to each intent by the two genders ordered by the difference between the genders' responses (measured by the Kolmogorov-Smirnov (KS) p-value). The higher the p-value, the more similar the responses from male and female participants are. There are broadly similar priorities reported for most intents across the two genders. For instance, both genders frequently report using Tinder for finding a partner. More interesting is the intention that the two genders report profoundly different rankings on. The mode average score given by men to \textit{one night stands} is \modemalehookup, compared to just \modefemalehookup~by women (this polarisation is reflected in an extremely low KS p-value). \pmalehookup \% of male respondents rated ``I use Tinder for one night stands'' as 4 or 5, compared to just \pfemalehookup\% of women. Tinder's inability to differentiate between the intentions of its users therefore inevitably leads to many matchings between parties who are looking for very different things. This is hinted at by the extremely low rate at which people report meeting up with their matches. \prealworldundereleven\% of respondents estimated that 10\% or fewer of their matches result in a real-world meet-up --- as one would expect, the distribution of meet-up estimates are near identical across both male and female respondents (for heterosexual users, these are interdependent).

\begin{table}[t]
\center
\begin{tabular}{l | c | c | c }
\hline
& Men & Women & KS \\
\hline\hline
One night stand & 3.2 & 1.8 & 0.00034 \\
Chat & 3.0 & 3.7 & 0.29122 \\
Casual Dating & 3.8 & 3.3 & 0.33535 \\
Look at Profiles & 3.5 & 3.8 & 0.35559 \\
Meet Partner & 3.5 & 3.3 & 0.99094 \\
\hline
\end{tabular}
\caption{Mean score for each intent when using Tinder (ordered by the empirical distance between male and female answers)}
\label{tab:survey:intent}
\end{table}

To investigate how these intentions translate to liking strategies, we next asked participants \textit{``Which liking strategies do you most frequently employ?''}. Multiple options could be selected from:
\one~I casually like most profiles; 
\two~I only like profiles that I'm attracted to; 
\three~I adapt my selectivity based on how many matches I am getting that day; 
\four~I adapt my selectivity based on what I am looking for at the time (\eg chat, date, relationship); and
\five~Other.
Table~\ref{tab:survey:strategy} presents the results as the percentage of respondents stating that they use a given strategy frequently. Clear differences between the genders emerge. Most blatant is the ``I casually like most profiles'' strategy. Whereas \pmenwhocasuallylike\% of men report using this strategy regularly, no women report ever using this. Instead, \pwomenwhoonylikeattracted\% of women report exclusively liking profiles they are explicitly attracted to (this is how Tinder is intended to be used). Interestingly, \pmenwhoadapt\% of men also state that they regularly adapt their liking rate based on how many matches they are receiving.

\begin{figure}[t]
 \center
 \includegraphics[width=80mm]{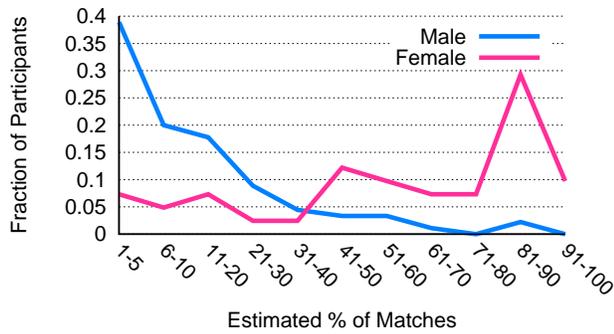}
 \caption{Probability density of estimated percentage of likes that result in a match}
  \label{fig:survey:estimate_match}
\end{figure}

We can now combine the above observations to explore how these liking strategies impact the matching rates of the opposite gender. To do this, we asked respondents to estimate the percentage of their likes that turn into matches. Figure~\ref{fig:survey:estimate_match} presents the results as a probability density function. In-line with our earlier empirical findings, women report far higher probabilities of matching, with \pwomenwhomatchoverfifty\% of women estimating that over half of the profiles they like result in a match. With such high matching rates, it is unsurprising that women use a strategy whereby they only like profiles they are confidently attracted to. In contrast, \pmenmatchundereleven\% of male respondents estimate that 10\% or under of their likes result in a match. A possible interpretation of the above data is that women get higher matching rates simply because they put more effort in, and find better potential partners to like. However, the very low messaging rates indicate that this is not the case: \pwomenchatunderthirtymatches\%~of women estimate that under 30\% of their matches result in a conversation.

These findings support our earlier suspicions, and partially explain why our female profiles found it easier to obtain high matching rates: a notable proportion of men \textit{casually} like most profiles. \pmencasuallikerswholikingoverfifty\% of male respondents who use this strategy report liking in excess of half of all women encountered. Some male respondents explicitly stated that this was caused by the rarity of matches they achieve, adding weight to our earlier discussion of an intuitive ``feedback loop''. This behaviour not only undermines the functionality of Tinder, but also shows that, contrary to intuition, Tinder cannot necessarily be used as an accurate tool for measuring (female) attractiveness or certain social phenomena (\eg homophily).

\begin{table}[t]
\center
\begin{tabular}{p{54mm} | c | c  }
\hline
 & Men & Women  \\
\hline\hline
I casually like most profiles & 35\% &  0\%  \\
I only like profiles that I'm attracted to & 72\% & 91\% \\
I adapt my selectivity based on how many matches I am getting that day & 13\% &  4\% \\
I adapt my selectivity based on what I am looking for at the time & 16\% & 22\% \\
\hline
\end{tabular}
\caption{Percentage of participants who reported frequently using each strategy (ordered by the empirical distance between male and female answers)}
\label{tab:survey:strategy}
\end{table}

\section{Conclusion} \label{sec:conclusion}

We have presented a study of user activity in Tinder, exploring how it differs between the two genders. We began by asking \one~what impact does gender have on matching and messaging rates, and \two~what profile characteristics are common across the two genders? We have shown that male users like a far higher proportion of profiles than females. Women, however, have a greater propensity to establish conversation via messaging, although they tend to leave a longer interval between matching and messaging than male users. It therefore seems that, rather than pre-filtering their mates via the like feature, many male users like in a relatively non-selective way and post-filter after a match has been obtained. This gaming of the system undermines its operation and likely leads to much frustration. Through controlled experiments, we have also observed clear attributes that can increase a user's ability to match, \eg bios and profile pictures. This is something that can help inform Tinder users in how to improve their online matching rates. 


It is important to note that our measurement study has limitations. Most notably, we only sample users from the vantage of a relatively small set of accounts in two urban areas. These do not necessarily cover the whole spectrum of attractiveness and, therefore, they may miss out matches that would have happened otherwise. So far, we have also treated Tinder as a black-box; the exact mechanism by which it selects profiles to present is unknown. Reverse-engineering and improving this could be highly productive and could add extra depth to our results (although note that all our profiles were setup in the same manner, ensuring equivalent comparisons). Beyond this, there is a wealth of exciting lines of future work that remain. Most important is expanding to more sophisticated setups, \eg comparing profiles based on physical characteristics such as height or race. Through this, we are keen to further explore our hypotheses and confirm that our observations are generalisable across more profile types. 




\footnotesize

\bibliographystyle{ieeetr}
\vspace{0.5cm}

{\balance\bibliography{Tinder}}

\end{document}

%% file: stats.tex
\def\nprefilteredrespondents{174\xspace}
\def\nresponses{131\xspace}
\def\nmales{90\xspace}
\def\nfemales{41\xspace}
\def\ponlinedatedbefore{68}
\def\puseonceaweek{87}
\def\pwomenlikeunder11{59}
\def\pmenlikeunder11{9}
\def\modemalehookup{5}
\def\modefemalehookup{1}
\def\pmalehookup{49}
\def\pfemalehookup{15}
\def\pmenlookingforhookupslikingover50{50}
\def\pmenwhocasuallylike{33}
\def\pwomenwhocasuallylike{0}
\def\pmenwhoonylikeattracted {70}
\def\pwomenwhoonylikeattracted{93}
\def\pmenwhoadapt {13}
\def\pmenwhomatchoverseventy {2}
\def\pwomenwhomatchoverseventy {46}
\def\pmenwhomatchoverfifty {7}
\def\pwomenwhomatchoverfifty {63}
\def\pmenmatchundereleven {59}
\def\meanrankmalebio {3}
\def\meanrankfemalebio {4}
\def\moderankfemalebio {4}
\def\pwomenchatunderthirtymatches {49}
\def\phetero {98}
\def\pmenlikeoverninety {10}
\def\pwomenlikeoverninety {0}
\def\pmencasuallikerswholikingoverfifty {80}
\def\prealworldundereleven {73}
\def\punderthirtyfive {83}